\documentstyle[prb,aps,multicol,amsmath]{revtex}
\def\eins{\mbox{\rm 1\hspace{-0.25em}l}}

\begin{document}
\draft
\title{Optical response and spin relaxation in semiconductor systems under 
excitation with arbitrary polarization}
\author{U. R\"{o}ssler}
\address{Institut f\"{u}r Theoretische Physik, Universit\"{a}t Regensburg, 
        {\em D-93040} Regensburg}
\address{and}
\address{Departamento de F\'isica de Materiales, Universidad Autonoma, Cantoblanco,
        {\em E-28049} Madrid}  
\date{\today}
\maketitle

\bigskip
\centerline{(accepted for publication in Phys. Stat. Sol. (b) {\bf 234} (2002))}

\begin{abstract}
\\ \\ \\ 
The equations-of-motion for the density matrix are derived in a
multiband model to describe the response of semiconductors (bulk or
quantum well structures) under optical excitation with arbitrary polarization. 
The multiband model used, comprising the twofold conduction
band and the fourfold topmost valence band (or heavy- and light-hole states),
incorporates spin-splitting of the single-particle states. 
The interaction terms include besides the direct Coulomb coupling between carriers 
also the electron-hole exchange interaction, which together with the spin-splitting 
terms is responsible for spin relaxation. Applying the Hartree-Fock truncation 
scheme leads
to a set of coherent semiconductor Bloch equations for the multiband case.
This concept provides the theoretical frame for describing phenomena connected with optical
response under excitation with arbitrary light polarization and spin relaxation: 
polarized optical response, polarization dynamics of VCSELs, spin relaxation, and 
the circular photovoltaic effect. \\ 
\end{abstract}
\pacs{PACS Num.~72.25.Fe,~72.25.Rb,~78.47.+p}


\newpage

{\large\section{INTRODUCTION}}               

The spin degree-of-freedom of electrons has recently attracted increasing attention 
in the context of {\em spintronics}.\cite{spintronics} In semiconductor quantum 
structures it is intimately connected with circularly
polarized light that is frequently used to create spin-polarized carriers
and to detect their lifetime with respect
to spin relaxation in pump-probe experiments.\cite{Damen,Heberle,Ohno} 
The investigation of optical orientation and of the mechanisms
of spin relaxation, first carried out in bulk semiconductors\cite{Piktit}, have become 
the basis for ongoing studies in semiconductor quantum
wells (QW), which are the likely material structure to realize the
spin transistor.\cite{Datta} Theoretical work devoted to spin relaxation makes
use preferentially of the density matrix to describe the thermal expectation
value of the spin as an observable.\cite{DP} For spin-polarized electrons
in the conduction band of a bulk semiconductor or in the lowest electron subband
of a QW structure it is sufficient to have the $2\times 2$
spin-density matrix and its time evolution due to generation/annihilation and
scattering processes. However, an adequate modelling of polarized photoluminescence involving
interband transitions, {\it i.e.} the calculation of the thermal expectation value of
the dipole operator for circularly polarized light, requires the knowledge of
the density matrix for conduction- and valence-band states in
a multiband scheme.\cite{Binder95} The dependence on the intensity of the exciting 
light is an important experimental aspect and should be included in the density 
matrix.\cite{Khitrova}

For vertical-cavity surface-emitting lasers (VCSEL) the phenomenon of
polarization instability has recently become a topic of
interest.\cite{SanMiguel,Ando} The VCSEL geometry allows for two independent
polarizations of the laser light, while VCSEL modelling, using the optical
or semiconductor Bloch equations (SBE), was based so far on the two-level system that
does not account for polarization degrees-of-freedom.\cite{SBE,Haug} This deficiency has been
removed partially by formulating a phenomenological set of optical (or Maxwell)
Bloch equations for a pair of two-level systems, that can emit separately light
with right or left circular polarization and which are coupled by spin-flip
processes.\cite{SanMiguel}

Semiconductor Bloch equations can be studied in different time regimes: the
coherent regime - when the interband polarization is in phase with the exciting
electromagnetic field - is characterized by
the Rabi oscillations, while the quasi-equilibrium situation is described by the optical
(linear or nonlinear) susceptibility. In extending the two-level system, showing
one Rabi frequency, to a three (or actually six) level system, comprising the
spin-degenerate conduction, heavy- and light-hole bands, Binder and
Lindberg\cite{Binder} formulated a set of equations for the optical polarization
functions and carrier distributions to describe coherent nonlinear response of
this more realistic quantum-well model under excitation with circularly
polarized light. This treatment includes band mixing due to heavy-light hole
coupling in the Luttinger Hamiltonian and many-body effects in a Hartree-Fock(HF) 
truncation with
single-particle self energies determined by the direct Coulomb interaction. This 
concept has been applied successfully in studies of the
intervalence band coherence due to coupled optical Stark shifts.\cite{Donovan} However,
spin-flip processes that become possible due to band structure effects or due to
electron-hole exchange (not in the scope of these studies) were not  
considered. 
\\
Several more recent papers reported on the observation of the circular
photogalvanic effect (CPGE) in $n$- and $p$-doped semiconductor quantum well
structures.\cite{Ganichev} This effect, originally predicted for bulk
semiconductors\cite{Ivchenko}, is based on converting the helicity of light into
a directed stationary current. It represents a nonlinear response of the system
to the intense exciting light which can be described by a third rank tensor. The
effect is caused by creating a nonequilibrium carrier distribution in the
spin-split subbands by excitation with circularly polarized light. The 
dependence of the effect on the light intensity, {\it i.e.} its saturation
behavior, bears information on the spin-relaxation and should allow to extract
the characteristic times.\cite{GanDan} A microscopic description of these phenomena 
requires the formulation of the nonlinear response under excitation with circular 
(in general elliptic) polarization.

Among the different mechanisms of spin relaxation \cite{Sham} those named after
D'yakonov and Perel' (DP)\cite{DP} and Bir, Aronov, and Pikus (BAP)\cite{BAP} seem to
dominate in semiconductor structures not containing magnetic impurities. The
DP mechanism is a single-particle effect. It is intimately connected with
spin-splitting of the electronic states caused by spin-orbit coupling and
inversion asymmetry. The BAP mechanism results from electron-hole exchange
scattering and depends on the carrier concentration resulting from doping or
intense optical excitation. Both mechanisms are of the motional-narrowing
type, {\it i.e.} the spin-relaxation rate is proportional to the momentum
scattering time.\cite{Maialle,Degani}
In experiments with circularly polarized light in the visible spectral
range electron-hole pairs are created (bipolar optical orientation) and spin relaxation 
takes place due to both
mechanisms, a situation which was not accounted for so far in theoretical work. 
The BAP mechanism can be avoided in far-infrared experiments with monopolar
spin orientation connected with transitions between electron (or hole) 
subbands.\cite{Ganichev,GanDan}

In order to provide a general theoretical frame for all these phenomena the 
equations-of-motion for the density matrix are formulated in a multiband model
with the spin and polarization degrees-of-freedom being taken into account.
The system Hamiltonian
\begin{equation}
H = H_0 + H'(t) + H_{Coul}
\end{equation}
consists of the single-particle part, $H_0 + H'(t)$, describing the involved electron
states and the interaction with the electromagnetic field, and the Coulomb
interaction $H_{Coul}$.
The multiband model comprises the twofold conduction, heavy- and light-hole bands (or the
respective lowest subbands in a QW structure) as in Ref.~\cite{Binder} 
but includes the mechanisms of spin-splitting as origin of 
the DP mechanism. In Section II the equations-of-motion will be formulated by
considering only the single-particle part of $H$ in order to introduce the
multiband model and to get familiar with the notation.
Many-body effects (due to free carriers introduced by doping or created by
optical excitation) are considered in Section III. Besides the direct Coulomb 
interaction between carriers also the electron-hole exchange coupling is taken
into account. It gives rise to spin-flip processes and 
represents the origin of the BAP mechanism of spin-relaxation. By following 
Refs.~\cite{SBE,Haug,Binder} the Hartree-Fock truncation is applied in Section IV  
to get a closed set of equations for the elements of the density matrix that describe
the coherent part of the problem. They turn out to be more general than those known 
from the literature \cite{Binder95,Binder} by taking into account both the 
polarization and spin 
degrees-of-freedom. In Section V these equations will be discussed with respect to 
simplifications (and including phenomenological damping) which allow to recover published 
results and to point out applications to new nonlinear response phenomena addressed here.

{\large\section{THE SINGLE-PARTICLE PART}}

The single-particle Hamiltonian $H_0$ is conveniently written in the second
quantized form
\begin{equation}
H_0 = \sum_{\alpha m_{\alpha} \vec{k}} \varepsilon^{\alpha}_{m_{\alpha}}(\vec{k})
a^\dagger_{\alpha m_{\alpha}}(\vec{k})a_{\alpha m_{\alpha}}(\vec{k})
\label{H00}\end{equation}
where $\alpha, m_\alpha, \vec{k}$ are the quantum numbers of a complete
set of eigenstates with energy $\varepsilon^{\alpha}_{m_{\alpha}}(\vec{k})$ and
$a^\dagger (a)$ are creation(annihilation) operators of the corresponding
states. For bulk semiconductors $\alpha$ denotes the energy bands, $m_\alpha$
the degeneracy within a given band, and $\vec{k}$ the three-dimensional (3D) wave
vector. In the following we
restrict the basis to the lowest conduction band ($\alpha=c$)
with $m_c=\pm1/2$ and the topmost valence band ($\alpha=v$) with
$m_v=\pm 3/2$ or $\pm 1/2$ for the heavy and light holes, respectively. The
notation applies as well to QW structures with the lowest electron
($\alpha=c, m_c=\pm1/2$), heavy ($\alpha=v, m_v=\pm3/2$), and light hole 
($\alpha=v, m_v=\pm1/2$) subbands and a 2D wave vector $\vec{k}$. (An extension to more subbands would require
book keeping of subband indices, which is avoided here.) The
single-particle energies $\varepsilon^{\alpha}_{m_{\alpha}}(\vec{k})$ are obtained
by diagonalizing a $2\times 2\,(4\times4)$ Hamiltonian for the conduction\,(valence)
band states, which
may include besides the diagonal free-particle kinetic energy also nonparabolic
corrections from a higher order $\vec{k}\cdot\vec{p}$ decoupling.\cite{k-p} This includes
terms resulting from bulk-inversion asymmetry or from the asymmetry of the
quantum well and lead to a removal of the spin-degeneracy of the (sub)bands.\cite{Wissinger}
Thus, the single-particle wave functions are expanded in a basis of band-edge ($\vec{k}=0$)
Bloch functions $u_{\alpha m'_{\alpha}}(\underline{x})$ ($\underline{x}$ comprising 
space and spin coordinates) which for the bulk semiconductor reads
\begin{equation}
\psi_{\alpha m_\alpha \vec{k}}(\underline{x}) = e^{i\vec{k}\cdot\vec{r}}\sum_{m'_{\alpha}}
C^\alpha_{m_\alpha m'_\alpha}(\vec{k})u_{\alpha m'_\alpha}(\underline{x})\,.
\label{expansion-3D}\end{equation}
The coefficients $C^\alpha_{m_\alpha m'_\alpha}(\vec{k})$ describe within a band
$\alpha$ the mixing of the band-edge states at finite $\vec{k}$; for the valence band 
this includes coupling between heavy- and light-hole states. Note, that the angular-momentum 
classification is exact only at $\vec{k}=0$, where $C^\alpha_{m_\alpha
m'_\alpha}(0)=\delta_{m_\alpha m'_\alpha}$. For QW structures (with $z$ being the
growth direction) the single-particle wave function takes the form 
\begin{equation}
\psi_{\alpha m_\alpha \vec{k}}(\underline{x}) = e^{i\vec{k}\cdot\vec{r}}\sum_{m'_{\alpha}}
\zeta^{(m_\alpha)}_{m'_\alpha}(\vec{k},z)u_{\alpha m'_\alpha}(\underline{x})\,.
\label{expansion-2D}\end{equation}
Here $\vec{k}=(k_x,k_y,0)$ is the in-plane wave vector, $\vec{r}=(x,y,0)$, and 
$\zeta^{(m_\alpha)}_{m'_\alpha}(\vec{k},z)$ are the subband functions including $\vec{k}$ 
dependent mixing, while for $\vec{k}\rightarrow 0$ we have
$\zeta^{(m_\alpha)}_{m'_\alpha}(\vec{k},z)\rightarrow\zeta^{(m_\alpha)}_{m'_\alpha}(0,z)\delta_{m_\alpha m'_\alpha}$. 

With this formulation of the single-particle states $H_0$ is block-diagonal
in the band indices $\alpha=c,v$ and it is convenient to replace 
$a_{cm_c}(\vec{k}) \rightarrow c_{m_c}(\vec{k})$ for electrons 
and $a_{vm_v}(\vec{k}) \rightarrow v^\dagger_{-m_v}(-\vec{k})$ for holes 
(adopting the picture of holes as time-reversed electron states) and
similar for the hermitian adjoint fermion operators. In this notation 
(using the Kramers degeneracy $\varepsilon^{\alpha}_{-m_{\alpha}}(-\vec{k})=
\varepsilon^{\alpha}_{m_{\alpha}}(\vec{k})$) $H_0$ (Eq.~(2))
reads
\begin{equation}
H_0 = \sum_{m_c \vec{k}} \varepsilon^c_{m_c}(\vec{k})
c^\dagger_{m_c}(\vec{k})c_{m_c}(\vec{k}) + 
\sum_{m_v \vec{k}} \varepsilon^v_{m_v}(\vec{k})
v_{m_v}(\vec{k})v^{\dagger}_{m_v}(\vec{k})\,.
\label{H0}\end{equation}
Likewise the interaction with the electromagnetic field can be written
\begin{equation}
H'(t) = -\sum_{m_cm_v\vec{k}}\left\{\vec{E}(t)\cdot
\vec{d}^{cv}_{m_cm_v}(\vec{k})c^\dagger_{m_c}(\vec{k})v^\dagger_{-m_v}(-\vec{k})
+h.c.\right\}
\end{equation}
with the matrix of the dipole operator 
\begin{equation}
\vec{d}^{cv}_{m_cm_v}(\vec{k}) =
e\sum_{m'_cm'_v}C^{c*}_{m_cm'_c}(\vec{k})\vec{R}_{m'_cm'_v}\tilde{C}^v_{m'_vm_v}(\vec{k})\,
\end{equation}
for the bulk case (here $\tilde{C}^v_{m'_vm_v}(\vec{k})$ denote the
expansion coefficients of the hole wave functions) and 
\begin{equation}
\vec{d}^{cv}_{m_cm_v}(\vec{k}) =
e\sum_{m'_cm'_v}\int\zeta^{(m_c)*}_{m'_c}(\vec{k},z)\vec{R}_{m'_cm'_v}\zeta^{(m_v)}_{m'_v}(\vec{k},z)dz\,.
\end{equation}
for the QW case. Here the real electric field vector is
$\vec{E}(t)=\vec{E}(\omega)e^{i\omega t} + \vec{E}^*(\omega)e^{-i\omega t}$ 
($\vec{E}^*(\omega) = \vec{E}(-\omega)$) and
$\vec{R}_{m'_cm'_v}$, the matrix element of $\vec{r}$, is taken between the 
band-edge Bloch functions $u_{cm'_c}(\vec{x})$ and $u_{vm'_v}(\vec{x})$ (which are
angular-momentum eigenfunctions). For light with arbitrary polarization 
propagating in the $z$-direction the complex field amplitudes can be written
$\vec{E}_\pm(\omega)=E_0(\omega)(\vec{e}_x+\vec{e}_ye^{\pm i\phi})/2$ where
the $\pm$-sign indicates (for $\phi\neq 0$) the helicity of light. For
$\phi=\pi/2$ this is the right and left circular polarization. The matrix
elements of $\vec{E}_\pm(\omega)\cdot\vec{R}_{m'_cm'_v}$ are given in the
following table $(R=<S|x|X>/\sqrt{2})$:

\bigskip
 
\begin{tabular}{l|cccc}
$m_c=+\frac{1}{2}$ & $-\frac{E_0R}{2}(1+ie^{\pm i\phi})$ & 0 & 
$\frac{E_0R}{2\sqrt{3}}(1-ie^{\pm i\phi})$ & 0 \\
$m_c=-\frac{1}{2}$ & 0 & $-\frac{E_0R}{2\sqrt{3}}(1+ie^{\pm i\phi})$ & 0 & 
$\frac{E_0R}{2}(1-ie^{\pm i\phi})$ \\ \hline 
$m_v=$ & $\frac{3}{2}$ & $\frac{1}{2}$ & $-\frac{1}{2}$ & $-\frac{3}{2}$  
\end{tabular}

\bigskip  

The single-particle Hamiltonian $H_0+H'(t)$ is a sum of contributions from
different wave vectors $\vec{k}$ at each of which we have a six-level system
according to the multiband model. In analogy with the 
two-level model of Refs.~\cite{SBE,Haug} one can now formulate the equations-of-motion
for the operators $c^{\dagger}_{m_c}c_{\bar{m}_c}, v_{-m_v}v^{\dagger}_{-\bar{m}_v},$
and $v_{-m_v}c_{\bar{m}_c}$ 
\begin{eqnarray}
\big(i\hbar\partial_t - \varepsilon^c_{m_c}(\vec{k}) & + &
\varepsilon^c_{\bar{m}_c}(\vec{k})\big)c^{\dagger}_{m_c}(\vec{k},t)c_{\bar{m}_c}(\vec{k},t) =
\nonumber\\
& = & \sum_{m_v}\left\{\vec{E}(t)\cdot\vec{d}^{cv}_{\bar{m}_cm_v}(\vec{k})c^{\dagger}_{m_c}(\vec{k},t)v^{\dagger}_{-m_v}(-\vec{k},t)
- h.c.\right\}
\label{c+c}\end{eqnarray}
\begin{eqnarray} 
\big(i\hbar\partial_t + \varepsilon^v_{-\bar{m}_v}(-\vec{k}) & - &
\varepsilon^v_{-m_v}(-\vec{k})\big)v_{-m_v}(-\vec{k},t)v^{\dagger}_{-\bar{m}_v}(-\vec{k},t)
= \nonumber\\ 
& = & - \sum_{m_c}\left\{\vec{E}(t)\cdot\vec{d}^{cv}_{m_cm_v}(\vec{k})c^{\dagger}_{m_c}(\vec{k},t)v^{\dagger}_{-\bar{m}_v}(-\vec{k},t)
- h.c.\right\}               
\label{vv+}\end{eqnarray}
\begin{eqnarray}
\big(i\hbar\partial_t + \varepsilon^c_{\bar{m}_c}(\vec{k}) -
\varepsilon^v_{-m_v}(-\vec{k})\big)v_{-m_v}(-\vec{k},t)c_{\bar{m}_c}(\vec{k},t) & = & \nonumber\\
= \sum_{m'_v}\vec{E}(t)\cdot\vec{d}^{cv}_{\bar{m}_cm'_v}(\vec{k})v_{-m_v}(-\vec{k},t)v^{\dagger}_{-m'_v}(-\vec{k},t)               
& - & \sum_{m'_c}\vec{E}(t)\cdot\vec{d}^{cv}_{m'_cm_v}(\vec{k})c^{\dagger}_{m'_c}(\vec{k},t)c_{\bar{m}_c}(\vec{k},t)\,.
\label{vc}\end{eqnarray}
These operators can be identified according to
\begin{eqnarray}\label{RHO}
\hat{\rho}^c_{m_c\bar{m}_c}(\vec{k},t) & = &
c^{\dagger}_{m_c}(\vec{k},t)c_{\bar{m}_c}(\vec{k},t)\, , \quad\quad\quad
m_c,\bar{m}_c=\pm 1/2 \\
\hat{\rho}^v_{m_v\bar{m}_v}(\vec{k},t) & = &
v_{-m_v}(-\vec{k},t)v^{\dagger}_{-\bar{m}_v}(-\vec{k},t)\, ,\,
m_v,\bar{m}_v=\pm 1/2,\pm 3/2 \\
\hat{P}^\dagger_{m_v\bar{m}_c}(\vec{k},t) & = & v_{-m_v}(-\vec{k},t)c_{\bar{m}_c}(\vec{k},t)
\end{eqnarray}
with the operators $\hat{\rho}^c$ $(1-\hat{\rho}^v)$ of the spin density matrix for
electrons (holes) and $\hat{P}$ of the interband polarization matrix. The matrix
$\vec{E}(t)\cdot\vec{d}^{cv}_{m_cm_v}(\vec{k})$ is the multiband generalization
to the (unrenormalized) Rabi frequency of the two-level model. When taking the thermal average of
the matrix operators Eqs.~(\ref{c+c})-(\ref{vc}) correspond to the single-particle
parts of Eqs.~(1)-(3) of Ref.~\cite{Binder}, which however do not include
spin splitting of the electron and hole eigenstates. (Note, that in the present
case the conduction- and valence-band Hamiltonians are diagonalized, see
Eq.~(\ref{H0})).

\bigskip

{\large\section{THE MANY-BODY TERMS}}

In correspondence with the formulation of the single-particle part of the system
Hamiltonian the Coulomb interaction
\begin{equation}
H_{Coul} = \frac{1}{2}\int d\underline{x}\int
d\underline{x}'\Psi^{\dagger}(\underline{x})\Psi^{\dagger}(\underline{x}')
v(|\vec{r}-\vec{r}'|)\Psi(\underline{x}')\Psi(\underline{x})
\label{HCoul}\end{equation}
is decomposed by splitting the field operator 
\begin{equation}
\Psi(\underline{x}) = \Psi_e(\underline{x}) + \Psi^{\dagger}_h(\underline{x})
\end{equation}
into its electron and hole parts, where $\underline{x}$ stands again 
for space and spin variables. Thus one obtains
\begin{equation}
H_{Coul} = H_{ee} + H_{hh} + H^C_{eh} + H^X_{eh}
\end{equation}
where the individual terms have the same form for the bulk and QW case (time-dependence
of the creation and annihilation operators is understood)
\begin{eqnarray}
H_{ee} & = &
\frac{1}{2}\sum_{m_{c_1}m_{c_2}\atop
m_{c_3}m_{c_4}}\sum_{\vec{k}\vec{k}'\atop\vec{q}}{\cal
V}^{ee}_{m_{c_1}m_{c_2}m_{c_3}m_{c_4}}(\vec{k},\vec{k}',\vec{q})
c^{\dagger}_{m_{c_1}}(\vec{k}+\vec{q})c^{\dagger}_{m_{c_2}}(\vec{k}'-\vec{q})
c_{m_{c_3}}(\vec{k'})c_{m_{c_4}}(\vec{k})\,, \\
H_{hh} & = &
\frac{1}{2}\sum_{m_{v_1}m_{v_2}\atop
m_{v_3}m_{v_4}}\sum_{\vec{k}\vec{k}'\atop\vec{q}}{\cal
V}^{hh}_{m_{v_1}m_{v_2}m_{v_3}m_{v_4}}(\vec{k},\vec{k}',\vec{q})
v_{m_{v_1}}(\vec{k}-\vec{q})v_{m_{v_2}}(\vec{k}'+\vec{q})
v^{\dagger}_{m_{v_3}}(\vec{k}')v^{\dagger}_{m_{v_4}}(\vec{k})\,, \\
H^C_{eh} & = &
-\sum_{m_{c_1}m_{c_2}\atop m_{v_1}m_{v_2}}\sum_{\vec{k}\vec{k}'\atop\vec{q}}{\cal
V}^{eh,C}_{m_{c_1}m_{v_1}m_{v_2}m_{c_2}}(\vec{k},\vec{k}',\vec{q})
c^{\dagger}_{m_{c_1}}(\vec{k}+\vec{q})c_{m_{c_2}}(\vec{k})
v^{\dagger}_{m_{v_2}}(\vec{k}')v_{m_{v_1}}(\vec{k}'+\vec{q})\,, \\
H^X_{eh} & = &
\sum_{m_{c_1}m_{c_2}\atop m_{v_1}m_{v_2}}\sum_{\vec{k}\vec{k}'\atop\vec{q}}{\cal
V}^{eh,X}_{m_{c_1}m_{v_1}m_{c_2}m_{v_2}}(\vec{k},\vec{k}',\vec{q})
c^{\dagger}_{m_{c_1}}(-\vec{k}+\vec{q})c_{m_{c_2}}(-\vec{k}'+\vec{q})
v^{\dagger}_{m_{v_2}}(\vec{k})v_{m_{v_1}}(\vec{k}')\,.
\end{eqnarray}
They describe the interaction between electrons in the conduction band ($H_{ee}$), 
between holes in the valence band ($H_{hh}$), and the direct electron-hole Coulomb 
interaction ($H^C_{eh}$). These terms are known from previous treatments of the 
problem.\cite{Khitrova,SBE,Binder} The electron-hole exchange interaction ($H^X_{eh}$),
however, has not been considered so far in the context of SBE.\cite{Kuklinski} 
Together with the
spin-splitting mechanism in the single-particle part (Section II) it is the essential 
extension of the existing theory and will be discussed in the course of this paper.

Due to the diagonalization of the (sub)band Hamiltonians for electrons and holes
(see Eqs.~(\ref{H00}) and
(\ref{expansion-2D})) the interaction matrixelements depend on the expansion in terms of the
band-edge Bloch functions. For the bulk system they read 
\begin{eqnarray}
{\cal V}^{ee}_{m_{c_1}...m_{c_4}}(\vec{k},\vec{k'},\vec{q}) & = & 
v(q)\sum_{m_cm'_c}C^{c*}_{m_{c_1}m_c}(\vec{k}+\vec{q})C^{c*}_{m_{c_2}m'_c}(\vec{k'}-\vec{q})
                         C^c_{m_{c_3}m'_c}(\vec{k'})C^c_{m_{c_4}m_c}(\vec{k})\,,\\
{\cal V}^{hh}_{m_{v_1}...m_{v_4}}(\vec{k},\vec{k'},\vec{q}) & = & 
v(q)\sum_{m_vm'_v}\tilde{C}^{v*}_{m_{v_1}m_v}(\vec{k}-\vec{q})\tilde{C}^{v*}_{m_{v_2}m'_v}(\vec{k'}+\vec{q})                        
		  \tilde{C}^v_{m_{v_3}m'_v}(\vec{k'})\tilde{C}^v_{m_{v_4}m_v}(\vec{k})\,,\\
{\cal V}^{eh,C}_{m_{c_1}...m_{c_2}}(\vec{k},\vec{k'},\vec{q}) & = & 
v(q)\sum_{m_cm_v}C^{c*}_{m_{c_1}m_c}(\vec{k}+\vec{q})C^c_{m_{c_2}m_c}(\vec{k})
                        \tilde{C}^{v*}_{m_{v_1}m_v}(\vec{k'}+\vec{q})\tilde{C}^v_{m_{v_2}m_v}(\vec{k'})\,,\\			 
{\cal V}^{eh,X}_{m_{c_1}...m_{v_2}}(\vec{k},\vec{k'},\vec{q}) & = & 
v(q)\sum_{m_cm_vm'_cm'_v}{\cal
M}^{m_cm'_v}_{m_vm'_c}C^{c*}_{m_{c_1}m_c}(-\vec{k}+\vec{q})C^c_{m_{c_2}m'_c}(-\vec{k'}+\vec{q})\times \nonumber\\
           & & \mbox{\hspace{7cm}}\times\tilde{C}^{v*}_{m_{v_1}m_v}(\vec{k'})\tilde{C}^v_{m_{v_2}m'_v}(\vec{k})\,
\end{eqnarray}
with the expansion coefficients $C^c$ for electron and $\tilde{C}^v$ for
hole wavefunctions. Here $v(q)$ is the
Fourier transformed 3D Coulomb potential. For the QW case one has instead
\begin{eqnarray}
{\cal V}^{ee}_{m_{c_1}m_{c_2}m_{c_3}m_{c_4}}(\vec{k},\vec{k'},\vec{q}) & = & 
v(q)\sum_{m_cm'_c}\int dz\int dz'e^{-q|z-z'|}\times\nonumber\\
& \times & \zeta^{(m_{c_1})*}_{m_c}(\vec{k}+\vec{q},z)\zeta^{(m_{c_2})*}_{m'_c}(\vec{k'}-\vec{q},z')
                         \zeta^{(m_{c_3})}_{m'_c}(\vec{k'},z')\zeta^{(m_{c_4})}_{m_c}(\vec{k},z)
\end{eqnarray}
and corresponding expressions for ${\cal V}^{hh}, {\cal V}^{eh,C},$ and ${\cal V}^{eh,X}$, with 
the band-mixing at finite in-plane $\vec{k}$ considered in the subband functions. Here
$v(q)$ is the 2D Coulomb potential (see {\it e.g.} \cite{Khitrova} Eq.\,(A.18)). 
The Matrix ${\cal M}$ in the exchange interaction is due to the coupling of 
electron and hole angular momenta, known from studies of the finestructure splitting 
of excitons\cite{Denisov,Trebin,Jorda} or of the BAP spin-relaxation mechanism\cite{Maialle,Degani}. 
Without mixing of the band-edge states the interaction
matrixelements ${\cal V}^{ee}, {\cal V}^{hh}$, and ${\cal V}^{eh,C}$ take much 
simpler forms as {\it e.g.} ${\cal
V}^{eh,C}_{m_{c_1}m_{v_1}m_{v_2}m_{c_2}}(\vec{k},\vec{k'},\vec{q}) 
\rightarrow v(q)\delta_{m_{c_1}m_{c_2}}\delta_{m_{v_1}m_{v_2}}$ and
${\cal V}^{eh,X}_{m_{c_1}m_{v_1}m_{c_2}m_{v_2}} \rightarrow 
v(q){\cal M}^{m_{c_1}m_{v_ 2}}_{m_{v_1}m_{c_2}}$ for the 3D (or bulk) case and similar for 
the QW case. 

By calculating the commutators of the operators from Eqs.~(12)-(14) with
$H_{Coul}$ one finds the following additional terms which are to be added to the
Eqs.\,(\ref{c+c})-(\ref{vc}):
\begin{eqnarray}\label{CCH}
\Big[H_{Coul} & , & c^{\dagger}_{m_c}(\vec{k})c_{\bar{m}_c}(\vec{k})\Big] = \nonumber\\
= & - & \sum_{\vec{\bar{k}}\vec{q}m'_c}\Big\{\sum_{m_{c_1}m_{c_4}}
\Big( {\cal V}^{ee}_{m_{c_1}\bar{m}_cm'_cm_{c_4}}(\vec{\bar{k}},\vec{k}+\vec{q},\vec{q})
c^{\dagger}_{m_{c_1}}(\vec{\bar{k}}+\vec{q})c^{\dagger}_{m_c}(\vec{k})c_{m'_c}(\vec{k}+\vec{q})c_{m_{c_4}}(\vec{k})
\nonumber\\
& - & {\cal V}^{ee}_{m_{c_1}m'_cm_cm_{c_4}}(\vec{\bar{k}},\vec{k},\vec{q})
c^{\dagger}_{m_{c_1}}(\vec{\bar{k}}+\vec{q})c^{\dagger}_{m'_c}(\vec{k}-\vec{q})c_{\bar{m}_c}(\vec{k})c_{m_{c_4}}(\vec{\bar{k}})
\Big) \nonumber\\ 
& - & \sum_{m_{v_1}m_{v_2}}\Big({\cal V}^{eh,C}_{\bar{m}_cm_{v_1}m_{v_2}m'_c}(\vec{k}-\vec{q},\vec{\bar{k}},\vec{q})
c^{\dagger}_{m_c}(\vec{k})c_{m'_c}(\vec{k}-\vec{q})  \nonumber\\
& - & {\cal V}^{eh,C}_{m'_cm_{v_1}m_{v_2}m_c}(\vec{k},\vec{\bar{k}},\vec{q})
c^{\dagger}_{m'_c}(\vec{k}+\vec{q})c_{\bar{m}_c}(\vec{k})\Big)v^{\dagger}_{m_{v_2}}(\vec{\bar{k}})v_{m_{v_1}}(\vec{\bar{k}}+\vec{q})
\nonumber\\
& + & \sum_{m_{v_1}m_{v_2}}\Big(
{\cal V}^{eh,X}_{\bar{m}_cm_{v_1}m'_cm_{v_2}}(-\vec{k}+\vec{q},\vec{\bar{k}},\vec{q})
c^{\dagger}_{m_c}(\vec{k})c_{m'_c}(-\vec{\bar{k}}+\vec{q})v^{\dagger}_{m_{v_2}}(-\vec{k}+\vec{q})v_{m_{v_1}}(\vec{\bar{k}})
\nonumber\\
& - & {\cal V}^{eh,X}_{m'_cm_{v_1}m_cm_{v_2}}(\vec{\bar{k}},-\vec{k}+\vec{q},\vec{q})
c^{\dagger}_{m'_c}(-\vec{\bar{k}}+\vec{q})c_{\bar{m}_c}(\vec{k})v^{\dagger}_{m_{v_2}}(\vec{\bar{k}})v_{m_{v_1}}(-\vec{k}+\vec{q})
\Big)\Big\},
\end{eqnarray}
\begin{eqnarray}\label{VVH}
\Big[H_{Coul} & , & v_{-m_v}(-\vec{k})v^{\dagger}_{-\bar{m}_v}(-\vec{k})\Big] = \nonumber\\
= & - & \sum_{\vec{\bar{k}}\vec{q}m'_v}\Big\{\sum_{m_{v_1}m_{v_4}}\Big( 
{\cal V}^{hh}_{m_{v_1}-\bar{m}_vm'_vm_{v_4}}(\vec{\bar{k}},-\vec{k}-\vec{q},\vec{q})
v_{m_{v_1}}(\vec{\bar{k}}-\vec{q})v_{-m_v}(-\vec{k})v^{\dagger}_{m'_v}(-\vec{k}-\vec{q})v^{\dagger}_{m_{v_4}}(\vec{\bar{k}})
\nonumber\\
& - & {\cal V}^{hh}_{m_{v_1}m'_v-m_vm_{v_4}}(\vec{\bar{k}},-\vec{k},\vec{q})
v_{m_{v_1}}(\vec{\bar{k}}-\vec{q})v_{m'_v}(-\vec{k}+\vec{q})v^{\dagger}_{-\bar{m}_v}(-\vec{k})v^{\dagger}_{m_{v_4}}(\vec{\bar{k}})
\Big)\nonumber\\
& - & \sum_{m_{c_1}m_{c_2}}\Big({\cal V}^{eh,C}_{m_{c_1}m'_v-m_vm_{c_2}}(\vec{\bar{k}},-\vec{k},\vec{q})
v_{m'_v}(-\vec{k}+\vec{q})v^{\dagger}_{-{\bar{m}_v}}(-\vec{k}) \nonumber\\
& - & {\cal V}^{eh,C}_{m_{c_1}-\bar{m}_vm'_vm_{c_2}}(\vec{\bar{k}},-\vec{k}-\vec{q},\vec{q})
v_{-m_v}(-\vec{k})v^{\dagger}_{m'_v}(-\vec{k}-\vec{q})\Big)c^{\dagger}_{m_{c_1}}(\vec{\bar{k}}+\vec{q})c_{m_{c_2}}(\vec{\bar{k}})
\nonumber\\
& + & \sum_{m_{c_1}m_{c_2}}\Big( 
{\cal V}^{eh,X}_{m_{c_1}m'_vm_{c_2}-m_v}(-\vec{k},\vec{\bar{k}},\vec{q})
c^{\dagger}_{m_{c_1}}(\vec{k}+\vec{q})c_{m_{c_2}}(-\vec{\bar{k}}+\vec{q})v_{m'_v}(\vec{\bar{k}})v^{\dagger}_{-\bar{m}_v}(-\vec{k})
\nonumber\\
& - & {\cal V}^{eh,X}_{m_{c_1}-\bar{m}_vm_{c_2}m'_v}(\vec{\bar{k}},-\vec{k},\vec{q})
c^{\dagger}_{m_{c_1}}(-\vec{\bar{k}}+\vec{q})c_{m_{c_2}}(\vec{k}+\vec{q})v_{-m_v}(-\vec{k})v^{\dagger}_{m'_v}(\vec{\bar{k}})
\Big)\Big\},
\end{eqnarray}
\begin{eqnarray}\label{VCH}
\Big[H_{Coul} & , & v_{-m_v}(-\vec{k})c_{\bar{m}_c}(\vec{k})\Big] = \nonumber\\
= & - & \sum_{m_{c_1}m_{c_3}m_{c_4}\atop\vec{\bar{k}}\vec{q}}
{\cal V}^{ee}_{m_{c_1}\bar{m}_cm_{c_3}m_{c_4}}(\vec{\bar{k}},\vec{k}+\vec{q},\vec{q})
v_{-m_v}(-\vec{k})c^{\dagger}_{m_{c_1}}(\vec{\bar{k}}+\vec{q})c_{m_{c_4}}(\vec{\bar{k}})c_{m_{c_3}}(\vec{k}+\vec{q})
\nonumber\\
& - & \sum_{m_{v_1}m_{v_2}m_{v_3}\atop\vec{\bar{k}}\vec{q}}{\cal V}^{hh}_{m_{v_2}m_{v_1}m_{v_3}-m_v}(-\vec{k},\vec{\bar{k}},-\vec{q})
v_{m_{v_1}}(\vec{\bar{k}}-\vec{q})v_{m_{v_2}}(-\vec{k}+\vec{q})v^{\dagger}_{m_{v_3}}(\vec{\bar{k}})c_{\bar{m}_c}(\vec{k})
\nonumber\\
& + & \sum_{m_{c_2}m_{v_1}\atop\vec{\bar{k}}\vec{q}}\Big(\sum_{m'_v}{\cal V}^{eh,C}_{\bar{m}_cm_{v_1}m'_vm_{c_2}}(\vec{k}-\vec{q},\vec{\bar{k}},\vec{q})
v_{-m_v}(-\vec{k})c_{m_{c_2}}(\vec{k}-\vec{q})v^{\dagger}_{m'_v}(\vec{\bar{k}})v_{m_{v_1}}(\vec{\bar{k}}+\vec{q}) \nonumber\\
& + & \sum_{m'_c}{\cal V}^{eh,C}_{m'_cm_{v_1}-m_vm_{c_2}}(\vec{\bar{k}},-\vec{k},\vec{q})
c^{\dagger}_{m'_c}(\vec{\bar{k}}+\vec{q})c_{m_{c_2}}(\vec{\bar{k}})v_{m_{v_1}}(-\vec{k}+\vec{q})c_{\bar{m}_c}(\vec{k})
\Big) \nonumber\\
& + & \sum_{m_{c_2}m_{v_1}\atop\vec{\bar{k}}\vec{q}}\Big(\sum_{m'_v}{\cal
V}^{eh,X}_{\bar{m}_cm_{v_1}m_{c_2}m'_v}(-\vec{k}+\vec{q},\vec{\bar{k}},\vec{q})
v_{-m_v}(-\vec{k})c_{m_{c_2}}(-\vec{\bar{k}}+\vec{q})v^{\dagger}_{m'_v}(-\vec{k}+\vec{q})v_{m_{v_1}}(\vec{\bar{k}}) \nonumber\\
& + &  \sum_{m'_c}{\cal V}^{eh,X}_{m'_cm_{v_1}m_{c_2}-m_v}(-\vec{k},\vec{\bar{k}},\vec{q})
c^{\dagger}_{m'_c}(\vec{k}+\vec{q})c_{m_{c_2}}(-\vec{\bar{k}}+\vec{q})v_{m_{v_1}}(\vec{\bar{k}})c_{\bar{m}_c}(\vec{k})
\Big).
\end{eqnarray}
Equations~(\ref{CCH}) and (\ref{VCH}) reduce to Eqs.~(A.29) and (A.28), respectively, of 
Ref.~\cite{Khitrova} when the band mixing and the exchange terms are neglected. 
The characteristic effect of the Coulomb interaction is to add four-operator terms to the 
equations of motion of the two-operator terms, thus giving rise to a hierarchy of such equations
which can be solved only approximately. The choice of the method will depend on the scenario that 
is to be described: the relevant time scale (to distinguish between the coherent or 
quasi-equilibrium regime), the carrier density, or the power of the driving electric 
field.\cite{SBE,Haug,Axt}.

\bigskip

{\large\section{THE COHERENT DENSITY-MATRIX EQUATIONS}}

A set of equations describing the coherent regime can be derived by following
Refs.~\cite{SBE,Haug} and applying the HF truncation to factorize the four-operator terms. 
By taking thermal expectation values and keeping only factors, which are
diagonal in the wave vector 
\begin {eqnarray}
c^{\dagger}_{m_c }(\vec{k},t)c_{\bar{m}_c}(\vec{k},t) & \rightarrow &
\langle c^{\dagger}_{m_c}(\vec{k},t)c_{\bar{m}_c}(\vec{k},t) \rangle = \rho^e_{m_c\bar{m}_c}(\vec{k},t)
\nonumber\\
v^{\dagger}_{-\bar{m}_v}(-\vec{k},t)v_{-m_v}(-\vec{k},t) & \rightarrow &
\langle v^{\dagger}_{-\bar{m}_v}(-\vec{k},t)v_{-m_v}(-\vec{k},t) \rangle =
\rho^h_{\bar{m}_vm_v}(\vec{k},t)
\nonumber\\
v_{-m_v}(-\vec{k},t)c_{\bar{m}_c}(\vec{k},t) & \rightarrow &
\langle v_{-m_v}(-\vec{k},t)c_{\bar{m}_c}(\vec{k},t) \rangle = P^*_{\bar{m}_cm_v}(\vec{k},t)
\end{eqnarray}
one finds a closed set of equations for the components of the density matrix
\begin{eqnarray}
\Big(i\hbar\partial_t - \varepsilon^c_{m_c}(\vec{k}) & + & 
\varepsilon^c_{\bar{m}_c}(\vec{k})\Big)\rho^e_{m_c\bar{m}_c}(\vec{k},t) =
\nonumber\\
= & + & \sum_{m_v}\Big\{\Omega^*_{\bar{m}_cm_v}(\vec{k},t)P_{m_cm_v}(\vec{k},t)
    -\Omega_{m_cm_v}(\vec{k},t)P^*_{\bar{m}_cm_v}(\vec{k},t)\Big\} \nonumber\\
& + & \sum_{m'_c}\Big\{\Sigma^c_{m_cm'_c}(\vec{k},t)\rho^e_{m'_c\bar{m}_c}(\vec{k},t)
                     - \rho^e_{m_cm'_c}(\vec{k},t)\Sigma^c_{m'_c\bar{m}_c}(\vec{k},t)\Big\}
\label{rhocc}\end{eqnarray}
\begin{eqnarray} 
\Big(i\hbar\partial_t - \varepsilon^v_{\bar{m}_v}(\vec{k}) & + & 
\varepsilon^v_{m_v}(\vec{k})\Big)\rho^h_{\bar{m}_vm_v}(\vec{k},t)
= \nonumber\\ 
= & - & \sum_{m_c}\Big\{\Omega_{m_c\bar{m}_v}(\vec{k},t)P^*_{m_cm_v}(\vec{k},t)
      -P_{m_c\bar{m}_v}(\vec{k},t)\Omega^*_{m_cm_v}(\vec{k},t)\Big\} \nonumber\\
& - & \sum_{m'_v}\Big\{\Sigma^v_{m_vm'_v}(\vec{k},t)\rho^h_{\bar{m}_vm'_v}(\vec{k},t)
                     - \rho^h_{m'_vm_v}(\vec{k},t)\Sigma^v_{m'_v\bar{m}_v}(\vec{k},t)\Big\}
\label{rhovv}\end{eqnarray}
\begin{eqnarray}
\Big(i\hbar\partial_t + \varepsilon^c_{\bar{m}_c}(\vec{k}) & - &
\varepsilon^v_{m_v}(\vec{k})\Big)P^*_{\bar{m}_cm_v}(\vec{k},t) = \nonumber\\
= & + &
\sum_{m'_v}\Omega^*_{\bar{m}_cm'_v}(\vec{k},t)(\delta_{m_vm'_v}-\rho^h_{m'_vm_v}(\vec{k},t))               
-\sum_{m'_c}\Omega^*_{m'_cm_v}(\vec{k},t)\rho^e_{m'_c\bar{m}_c}(\vec{k},t) \nonumber\\
& - & \sum_{m'_c}P^*_{m'_cm_v}(\vec{k},t)\Sigma^c_{m'_c\bar{m}_c}(\vec{k},t)
  +   \sum_{m'_v}\Sigma^v_{m_vm'_v}(\vec{k},t)P^*_{\bar{m}_cm'_v}(\vec{k},t)\,.
\label{PVC}\end{eqnarray}
These equations are generalizations of Eqs.~(1)-(3) in Ref.~\cite{Binder} in that they
include the spin-splitting mechanisms in the single-particle part and by considering also the
electron-hole exchange interaction in the electron and hole self energies:
\begin{eqnarray}
\Sigma^c_{m_cm'_c}(\vec{k},t) = -\sum_{\vec{q}} & \Big\{ & \sum_{\tilde{m}_c\tilde{m}'_c}
{\cal
V}^{ee}_{m'_c\tilde{m}_cm_c\tilde{m}'_c}(\vec{k}-\vec{q},\vec{k},\vec{q})\rho^e_{\tilde{m}_c\tilde{m}'_c}(\vec{k}-\vec{q},t)
\nonumber\\
& - & \sum_{m_vm'_v}{\cal V}^{eh,X}_{m'_c-m_vm_c-m'_v}(-\vec{k}+\vec{q},-\vec{k}+\vec{q},\vec{q})
\rho^h_{m_vm'_v}(\vec{k}-\vec{q},t)\Big\} \label{Sigmac}\\
\Sigma^v_{m_vm'_v}(\vec{k},t) = -\sum_{\vec{q}} & \Big\{ & \sum_{\tilde{m}_v\tilde{m}'_v}
{\cal
V}^{hh}_{-m'_v-\tilde{m}_v-m_v-\tilde{m}'_v}(-\vec{k}+\vec{q},-\vec{k},\vec{q})
\big(\delta_{\tilde{m}_v\tilde{m}'_v} -
\rho^h_{\tilde{m}'_v\tilde{m}_v}(\vec{k}-\vec{q},t)\big)
\nonumber\\
& + & \sum_{m_cm'_c}{\cal V}^{eh,X}_{m_c-m'_vm'_c-m_v}(-\vec{k},-\vec{k},\vec{q})
\rho^e_{m_cm'_c}(\vec{k}+\vec{q},t)\Big\} \label{Sigmav}\,.
\end{eqnarray}
The renormalized dipole matrix reads
\begin{equation}
\Omega^*_{m_cm_v}(\vec{k},t)=\vec{E}(t)\cdot\vec{d}^{cv}_{m_cm_v}(\vec{k}) + 
\sum_{m'_cm'_v\atop\vec{q}}{\cal
V}^{eh,C}_{m_c-m'_v-m_vm'_c}(\vec{k}-\vec{q},-\vec{k},\vec{q})P^*_{m'_cm'_v}(\vec{k}-\vec{q},t)
\end{equation}
and differs from the corresponding expression in Ref.~\cite{Binder} (Eq.~(6)) only by the alternative 
choice of the basis. 

The self-energy terms have to be evaluated by summing over pairs of angular momenta for
electrons and holes and thus couple to all components of the electron or hole density-matrix. 
Take, {\it e.g.}, the equation of motion for $\rho^e_{m_c\bar{m}_c}(\vec{k},t)$:
the diagonal terms with $\Sigma^c_{m_cm_c}$ and $\Sigma^c_{\bar{m}_c\bar{m}_c}$ 
renormalize the single-particle energies, while the off-diagonal terms 
correspond to spin-flip scattering. The latter can be due to
electron-electron interaction modified by 
band-mixing but also to electron-hole exchange interaction. In fact, by neglecting in the 
self-energy, Eq.~(\ref{Sigmac}), the band-mixing it can be seen that the electron-electron 
interaction (because of
${\cal V}^{ee}_{m'_c\tilde{m}_cm_c\tilde{m}'_c}\sim\delta_{m'_c\tilde{m}'_c}\delta_{m_c\tilde{m}_c}$)
does not contribute to spin flips while the second term (from electron-hole exchange
interaction) retains the sum over $m_v,m'_v$ and couples electron and hole angular momenta. 
As it flips an electron spin simultaneous with a hole spin (if holes are
present), it can be identified as the origin of the BAP mechanism. 
The self-energy corrections due to ${\cal V}^{ee}$ and ${\cal V}^{hh}$ contain the spin-scattering processes
mediated by band-mixing in combination with spin-orbit coupling. They are more efficient for holes
(due to their $p$-character), while spin-orbit coupling in the electron states (sometimes called
Rashba \cite{Rashba} and Dresselhaus term \cite{Dress}) results from band-mixing at finite $\vec{k}$. 
Thus Eqs.\,(\ref{rhocc})-(\ref{PVC}) represent an extension of the coherent SBE\cite{SBE,Haug} to the 
multiband case and arbitrary polarization of the driving electromagnetic field, which goes beyond 
Ref.\,\cite{Binder} by including the spin-flip processes. These equations provide the theoretical background 
to describe experiments in the coherent regime that address the spin-degree of freedom like spin-echo
or four-wave mixing experiments with elliptically polarized light (in the latter case one could go beyond 
the HF truncation by using the dynamically controlled truncation scheme\cite{Axt}).

In closing this section a compact form of the Eqs.~(31)-(33) is presented which becomes possible by 
making use of matrix notation.\cite{ICPS-26} In the sixfold space of the multiband model used here the 
Hamiltonian ${\mathbf{H}}_0(\vec{k})+\boldsymbol{\Sigma}(\vec{k})$ (without the
coupling to the electromagnetic field) is block diagonal with $2\times 2$ and $4\times 4$ diagonal
blocks for the single particle and self-energy parts of conduction
$\big({\mathbf{H}}^c_0(\vec{k})+{\boldsymbol{\Sigma}}^c(\vec{k})\big)$ and
valence band states $\big({\mathbf{H}}^v_0(\vec{k})+{\boldsymbol{\Sigma}}^v(\vec{k})\big)$,
respectively. The interaction with the electromagnetic field, described by the
matrix $-{\boldsymbol{\Omega}}(\vec{k},t)$ of the (renormalized) dipole
interaction, couples between the valence and conduction band states and has
off-diagonal blocks only. Similarly, the $6\times 6$ density matrix
${\boldsymbol{\rho}}(\vec{k},t)$ consists
of diagonal blocks, the $2\times 2$ electron density 
matrix $\boldsymbol{\rho}^e$ and the $4\times 4$ hole density matrix
$\eins - \boldsymbol{\rho}^h$, and
off-diagonal blocks for the interband polarization $\mathbf{P}$ (the upper right
$2\times 4$ block) 
and its hermitian conjugate. Written with these matrices Eqs.~(31)-(33) take the compact 
form of the Liouville-von Neumann equation
\begin{equation}
i\hbar\partial_t{\boldsymbol{\rho}}(\vec{k},t) =
[{\mathbf{H}_0}(\vec{k})+{\boldsymbol{\Sigma}}(\vec{k},t)-{\boldsymbol{\Omega}}(\vec{k},t),
{\boldsymbol{\rho}}(\vec{k},t)]
\end{equation}
where $[..,..]$ indicates the antisymmetrized matrix product.     

\bigskip

{\large\section{DISCUSSION}}

This Section is devoted to the demonstration that
Eqs.\,(\ref{rhocc})-(\ref{PVC}) provide the 
theoretical frame for describing different scenarios of current interest which are related 
to excitation with arbitrary light polarization and spin relaxation. These scenarios will be
(i) the coupled two-level systems considered by San Miguel {\it et al.}\cite{SanMiguel} in 
order to treat the polarization dynamics in VCELs, (ii) the quasi-equilibrium limit leading 
to a generalized inhomogeneous equation for the interband polarization matrix and (for the 
low-density limit) to the multiband Wannier-exciton equation, (iii) spin relaxation of 
optically oriented electrons which at the same time are due to the DP and BAP mechanisms, and
(iv) the circular photogalvanic effect with interband excitation.

(i) Following San Miguel {\it et al.}\cite{SanMiguel} in their phenomenological treatment of 
polarization dynamics in VCSELs one has to consider only the two-level systems consisting of heavy hole
($m_v=\pm3/2$) and electron states ($m_c=\pm1/2$) which are dipole coupled by circularly 
polarized light. Thus Eqs.\,(\ref{rhocc})-(\ref{PVC}) are simplified by neglecting the 
light-hole states and the band mixing. For circular polarization the matrix of Rabi 
frequencies (without
renormalization) takes the form $\Omega_{m_cm_v}\sim\delta_{m_c\pm1/2}\delta_{m_v\pm3/2}$.
The diagonal terms of Eqs.\,(\ref{rhocc}) and (\ref{rhovv}) can be combined to write the 
equation-of-motion for the population inversion $D=\frac{1}{2}\{\rho^e_{\frac{1}{2}\frac{1}{2}}+
\rho^e_{-\frac{1}{2}-\frac{1}{2}} - 
(2-\rho^h_{\frac{3}{2}\frac{3}{2}}-\rho^h_{-\frac {3}{2}-\frac{3}{2}})\}$ as
\begin{equation}
\partial_t D = -\frac{1}{\tau_{||}}D +
\frac{2}{\hbar}\mbox{Im}(P_{\frac{1}{2}\frac{3}{2}}
\Omega^*_{\frac{1}{2}\frac{3}{2}} + P_{-\frac{1}{2}-\frac{3}{2}}\Omega^*_{-\frac{1}{2}-\frac{3}{2}})
\label{dtD}\end{equation}
where radiative decay is considered phenomenologically by the first term on the right 
hand side. This equation is (except for cavity effects, not considered here, and after adapting
the notation) identical with Eq.\,(2.5) of \cite{SanMiguel}. Likewise the equation-of-motion for 
the spin polarization $d=\frac{1}{2}\{\rho^e_{\frac{1}{2}\frac{1}{2}} + 
\rho^h_{\frac{3}{2}\frac{3}{2}} -
\rho^e_{-\frac{1}{2}-\frac{1}{2}} - \rho^h_{-\frac{3}{2}-\frac{3}{2}}\}$ is obtained in the form
\begin{eqnarray}
\partial_t d = -\frac{1}{\tau_{||}}d & + &
\frac{2}{\hbar}\mbox{Im}(P_{\frac{1}{2}\frac{3}{2}}\Omega^*_{\frac{1}{2}\frac{3}{2}} - 
P_{-\frac{1}{2}-\frac{3}{2}}\Omega^*_{-\frac{1}{2}-\frac{3}{2}}) \nonumber\\
& + & \frac{1}{i\hbar}\Big(\Sigma^c_{\frac{1}{2}-\frac{1}{2}}\rho^e_{-\frac{1}{2}\frac{1}{2}}-
\rho^e_{\frac{1}{2}-\frac{1}{2}}\Sigma^c_{-\frac{1}{2}\frac{1}{2}}
-\rho^h_{\frac{3}{2}-\frac{3}{2}}\Sigma^v_{\frac{3}{2}-\frac{3}{2}}+\Sigma^v_{-\frac{3}{2}\frac{3}{2}}\rho^h_{-\frac{3}{2}\frac{3}{2}}\Big)\,.
\label{dtd}\end{eqnarray}
In contrast to Eq.\,(\ref{dtD}), additional terms containing off-diagonal elements of the density and
self-energy matrices occur on the right hand side. When identifying these terms as $-\frac{2}{\tau_s}d$, 
describing spin relaxation \cite{taus}, one recovers Eq.\,(2.6) of
\cite{SanMiguel}. Finally Eq.\,(\ref{PVC}) can be 
formulated as
\begin{eqnarray}
\partial_tP_{\pm\frac{1}{2}\pm\frac{3}{2}} = & - & \Big(\frac{1}{\tau_\perp}+i\omega\Big)P_{\pm\frac{1}{2}\pm\frac{3}{2}}
+ \frac{i}{\hbar}\Omega_{\pm\frac{1}{2}\pm\frac{3}{2}}(D\pm d) \nonumber\\
& - &
\frac{i}{\hbar}\Big(\Sigma^c_{\pm\frac{1}{2}\mp\frac{1}{2}}P_{\mp\frac{1}{2}\pm\frac{3}{2}} 
- P_{\pm\frac{1}{2}\mp\frac{3}{2}}\Sigma^v_{\mp\frac{3}{2}\pm\frac{3}{2}}\Big)
\label{dtP}\end{eqnarray}
where $\hbar\omega=\varepsilon^c_{\pm\frac{1}{2}}(\vec{k})-\varepsilon^v_{\pm\frac{3}{2}}(\vec{k})$
and a phenomenological damping of the interband polarization was added. The last term (having a 
similar structure as that of Eq.\,(\ref{dtd})) can be understood as a contribution to this damping 
and combined with the first term to obtain Eq.\,(2.4) of \cite{SanMiguel}. Thus the phenomenological 
Maxwell-Bloch equations (especially those of Ref.\,\cite{SanMiguel} except for cavity effects) result 
as a special case of Eqs.\,(\ref{rhocc})-(\ref{PVC}) and give them a microscopic justification including
an interpretation of the spin-relaxation time $\tau_s$, which will become more transparent in the 
following.

(ii) Equations\,(\ref{rhocc})-(\ref{PVC}) in combination with
Eqs.\,(\ref{CCH})-(\ref{VCH}) involve different 
time-scales in the evolution of the density matrices for electrons and holes. Scattering due to 
terms neglected in the HF truncation lead on the sub-$ps$ time-scale to a loss of coherence 
concomitant with separate thermal equilibrium of electrons and holes (of
different spin) that decays by spin relaxation and electron-hole recombination.
This quasi-equilibrium can be described by replacing
\begin{equation}
\rho^e_{m_c\bar{m}_c}(\vec{k},t)\rightarrow f^c_{m_c}(\vec{k})\delta_{m_c\bar{m}_c},\quad 
\rho^h_{m_v\bar{m}_v}(\vec{k},t)\rightarrow
(1-f^v_{m_v}(\vec{k}))\delta_{m_v\bar{m}_v}
\end{equation}
where $f^c_{m_c}(\vec{k}), f^v_{m_v}(\vec{k})$ are the Fermi-Dirac distribution functions 
for conduction and valence band states, with quasi-Fermi levels depending on  
$m_c, m_v$. For this case the complex conjugate of Eq.\,(\ref{PVC}) may be reformulated 
by making partial use of the matrix notation (introduced
at the end of section IV): The matrix elements $P_{m_cm_v}$ form the
matrix of the interband polarization ${\mathbf{P}}(\vec{k},t)$,
$\big(\varepsilon^c_{m_c}(\vec{k})-\varepsilon^v_{m_v}(\vec{k})\big)P_{m_cm_v}(\vec{k},t)$
can be replaced by
$[{\mathbf{H}}_0(\vec{k}),{\mathbf{P}}(\vec{k},t)]_{m_cm_v}$, and 
the terms containing self energies by
$[{\boldsymbol{\Sigma}}(\vec{k},t),{\mathbf{P}}(\vec{k},t)]_{m_cm_v}$. 
Thus, Eq.\,(\ref{PVC}) takes the form
\begin{equation}
i\hbar\partial_tP_{m_cm_v}(\vec{k},t)-[{\mathbf{H}}_0(\vec{k}) + 
{\boldsymbol{\Sigma}}(\vec{k},t),{\mathbf{P}}(\vec{k},t)]_{m_cm_v} =
(f^c_{m_c}(\vec{k})-f^v_{m_v}(\vec{k}))\Omega_{m_cm_v}(\vec{k},t).
\label{RPA}\end{equation}
As in Eq.\,(\ref{dtP}) (to which it is related) one could add a phenomenological
damping term. Equation\,(\ref{RPA}) is the generalization of the RPA equation for the 
interband polarization (see Ref.\,\cite{Haug}) to the multiband case.

It is interesting here to simplify to the case without band-mixing, for which 
the electron-electron and hole-hole interactions become diagonal in the spin 
indices in contrast with the electron-hole exchange interaction. The latter contributes
to the off-diagonal elements of the electron self-energy matrix 
$\boldsymbol{\Sigma}^c(\vec{k})$ for nonvanishing hole concentration, thus 
indicating its relation to the BAP mechanism of spin relaxation.\cite{Degani}
Further specialization to the low-density limit leads to the multiband
generalization of the Wannier-exciton Schroedinger equation. 

(iii) Spin relaxation of electrons due to spin-orbit coupling combined with momentum scattering is 
described usually starting from an 
equation-of-motion for the electron spin-density matrix ${\boldsymbol{\rho}}^e(\vec{k},t)$.\cite{Piktit,DP} Such an 
equation is obtained from Eq.\,(\ref{rhocc}) by applying matrix notation as in Section IV but now only
for the $2\times 2$ electron block and identifying the first term on the {\it r.h.s.} as the generation 
matrix ${\mathbf{G}}(\vec{k},t)$ (see also Ref.\,\cite{Ivch-book})  
\begin{equation}
\partial_t{\boldsymbol{\rho}}^e(\vec{k},t) - \frac{1}{i\hbar}[{\mathbf{H}}^c_0(\vec{k})
+{\boldsymbol{\Sigma}}^c(\vec{k}),{\boldsymbol{\rho}}^e(\vec{k},t)]
+
\sum_{\vec{k}'}W(\vec{k},\vec{k}')\Big({\boldsymbol{\rho}}^e(\vec{k},t)-{\boldsymbol{\rho}}^e(\vec{k}',t)\Big) 
= {\mathbf{G}}(\vec{k},t)\,, 
\label{dtrhoe}\end{equation}
where we have added the last term on the {\it l.h.s.} to account for momentum scattering with
phonons and (nonmagnetic) impurities. Equation\,(\ref{dtrhoe}) is a generalized form of equations used 
in the context of spin relaxation due to the DP mechanism in that it combines the 
single-particle spin-relaxation mechanism due to spin-orbit coupling (DP) with the many-body 
mechanisms considered in the electron self energy ${\boldsymbol{\Sigma}}^c$, in particular the one caused 
by the electron-hole exchange interaction (BAP). To the best of our knowledge such a 
unified description of both mechanisms does not yet exist in the literature. A similar equation
following from Eq.\,(\ref{rhovv}) for the $4\times 4$ hole-density matrix 
${\boldsymbol{\rho}}^h(\vec{k},t)$ could be used to calculate the hole-spin relaxation.

(iv) Reported measurements of the photogalvanic effects (PGE) in QW structures have been performed with mid- and far-infrared
light on $p$- and $n$-doped samples.\cite{Ganichev,GanDan} The nonlinear excitation with elliptic 
polarization creates non-equilibrium populations of different spin states by intersubband transitions 
which in combination with scattering processes result in a stationary current density described by
\begin{equation}
j_{\alpha} = \chi_{\alpha\beta\gamma}E_{\beta}(\omega)E_{\gamma}(-\omega)  .
\end{equation}
It is ruled by a third rank tensor (similar to $\chi^{(2)}$ for second harmonic generation) that has 
nonvanishing components only in (so-called gyrotropic) systems lacking inversion symmetry. It can be
decomposed into contributions from linear and circular polarization. An 
alternative formulation of this stationary current density is given by\cite{Ivch-book}
\begin{equation}
j_{\alpha} = e{\mbox{Tr}}(\hat{\rho}\hat{v}_{\alpha})
\label{jalpha2}\end{equation}
where $\hat{v}_{\alpha}$ is a component of the velocity operator. As discussed in the literature the 
evaluation of Eq.\,(\ref{jalpha2}) requires to consider the off-diagonal elements of $\hat{\rho}$ and
$\hat{v}_{\alpha}$, where $\hat{\rho}$ contains the nonlinear dependence on the electric field 
amplitude of the exciting light. The theory developed here does not directly apply to this situation
but can be adapted to account for nonlinear intersubband excitation. On the other hand investigations 
of the CPGE with optical (or valence to conduction band) excitation, to which the concept presented 
here applies, are conceivable. An aspect of particular interest of PGE measurements is the dependence
of the saturation behavior on the light polarization, which as an alternative to time-resolved experiments
provides the possibility to detect the spin-relaxation time. Future work has to show how this dichroism
of the PGE saturation can be described in the frame of the concept presented here. 

\bigskip

{\bf Acknowledgment:} The work was performed at the Universidad Autonoma in
Madrid with support from the Ministerio de Educacion, 
Cultura y Deporte (Premio "A. von Humboldt - J.C. Mutis" 2000). Partial support came from the Volkswagen 
foundation and the Deutsche Forschungsgemeinschaft(DFG).

\bigskip
 
{\large\bf{REFERENCES}}


\begin{references}
\bibitem{taus} More precisely, one can identify 
$\frac{1}{i\hbar}(\Sigma^e_{\frac{1}{2}-\frac{1}{2}} \rho^e_{-\frac{1}{2}\frac{1}{2}} -
\rho^e_{\frac{1}{2}-\frac{1}{2}}\Sigma^e_{-\frac{1}{2}\frac{1}{2}})$ as 
$-\frac{2}{\tau_s}(\rho^e_{\frac{1}{2}\frac{1}{2}}-\rho^e_{-\frac{1}{2}-\frac{1}{2}})$
and similar for the hole contributions, with different spin-relaxation times for electrons and 
holes. In Ref.\,\cite{SanMiguel} $\tau^e_s=\tau^h_s$ has been explicitly assumed.
\end{references}
\end{document}